\documentclass[a4paper]{article}
\usepackage{authblk}
\usepackage{amsmath}
\usepackage{amsthm}
\usepackage{amsfonts}
\usepackage{amssymb}
\usepackage{mathtools}
\usepackage[round, numbers, sort]{natbib}
\usepackage{arxiv}

\usepackage{url}
\usepackage{hyperref}
\usepackage{indentfirst}

\usepackage{graphicx}
\usepackage{url}
\usepackage{orcidlink}

\begin{document}

\title{Warlock: an automated computational workflow for simulating spatially structured tumour evolution}

\date{}

\author[1]{Maciej Bak\,\orcidlink{0000-0003-1361-7301}}
\author[1]{Blair Colyer\,\orcidlink{0000-0002-9940-5707}}
\author[1]{Veselin Manojlovi\'{c}\,\orcidlink{0000-0003-0620-2431}}
\author[1]{Robert Noble*\,\orcidlink{0000-0002-8057-4252}}
\affil[1]{Department of Mathematics, City, University of London, London, UK}
\affil[*]{robert.noble@city.ac.uk}

\maketitle

\begin{abstract}
{A primary goal of modern cancer research is to characterize tumour growth and evolution, to improve clinical forecasting and individualized treatment. Agent-based models support this endeavour but existing models either oversimplify spatial structure or are mathematically intractable. Here we present \textit{warlock}, an open-source automated computational workflow for fast, efficient simulation of intratumour population genetics in any of a diverse set of spatial structures. \textit{Warlock} encapsulates a deme-based oncology model (\textit{demon}), designed to bridge the divide between agent-based simulations and analytical population genetics models, such as the spatial Moran process. Model output can be readily compared to multi-region and single-cell sequencing data for model selection or biological parameter inference. An interface for High Performance Computing permits hundreds of simulations to be run in parallel.
We discuss prior applications of this workflow to investigating human cancer evolution.}
	
\end{abstract}

\bigskip

\section*{Introduction}

Cancer evolution is a complicated dynamical system in which mutation, selection, genetic drift, and cell dispersal contribute to a patchwork of distinct cell subpopulations\cite{Greaves2012}. As these subpopulations can vary in terms of aggressiveness and treatment sensitivity, tumour heterogeneity is a promising subject of study in the emerging field of predictive oncology\cite{Dagogo-Jack2018, Maley2017, noble_when_2020}. Detailed characterization and understanding of tumour structure and evolutionary dynamics has potential to aid the design of precise, patient-specific prognoses and optimized targeted therapy regimens. Here we contribute to this effort by introducing the automated computational workflow \textit{warlock}, whose goal is to help scientists study the effects of spatial structure on tumour growth and evolution.

Several other agent-based modelling platforms exist that can be used to simulate tumour growth and evolution. \textit{PhysiCell}\cite{Ghaffarizadeh2018} specializes in off-lattice modelling of large numbers of cells in dynamic tissue microenvironments, with built-in resources for efficient diffusion field solving and dynamic cell-cycle state tracking. \textit{CompuCell3D}\cite{Swat2015} is capable of similar simulations, but with a different underlying model. \textit{HAL}\cite{Bravo2020} is a generic and highly customisable platform with modular components that allow for multiple grids working simultaneously to carry out different functions. \textit{Chaste}\cite{Mirams2013} allows the simulation of off- and on-lattice agent-based models and has been used for realistic simulations of defibrillation in human cardiac geometries and modelling colorectal crypts. \textit{J-SPACE}\cite{angaroni_j_2022} deals specifically with spatial models of cancer evolution and generates synthetic reads from Next-Generation Sequencing platforms.

The main contribution of our software is that it enables fast, efficient simulation of intratumour population genetics in any of a diverse set of biologically-relevant spatial structures. Our model’s unique deme-based structure, in which intra-deme dynamics resemble a Moran process, makes it especially appropriate for simulating the evolution of glandular tumours (each deme corresponding to a gland). This simple, generic structure also makes some aspects of the model amenable to mathematical analysis via diffusion approximations and other approaches. The aim is to bridge the divide between agent-based simulations of tumour evolution and tractable mathematical models of spatially structured population genetics. As such, our software is of potential use to a wide community of computational biologists, oncologists, population geneticists, and mathematical biologists.

\section*{Model overview}

At the heart of \textit{warlock} is a deme-based oncology model (\textit{demon}), a flexible framework for modelling intratumour population genetics with varied spatial structures and modes of cell dispersal. In this model, the spatial component is represented by a regular two-dimensional grid, where each point represents a \textit{deme} (a well-mixed patch of cells). We plan to support one- and three-dimensional grids in later versions. Cancer cells can divide, mutate, die, and move between demes, in manners that depend on the initial configuration of the simulation. The program tracks all driver and passenger mutations under the infinite sites assumption, supporting passenger mutation rates as high as one per cell division.

Input parameters determine initial cell division rates, mutation rates, mutation effects, cell dispersal modes (invasive or deme fission), dispersal rates, deme carrying capacity, spatial constraints, and stopping conditions. These parameters can be configured to simulate standard population genetics models such as the Eden growth model\cite{Eden}, voter model\cite{bramson_williams-bjerknes_1981}, spatial branching process\cite{Durrett2014a}, and spatial Moran process\cite{Durrett2014b}, as well as less conventional models. Optionally, the grid can be initially filled with normal cells to simulate the evolution of cancer as it invades normal tissue. Normal cells divide and die but neither mutate nor disperse.

\begin{figure}
\begin{center}
	\includegraphics[width=0.9\linewidth]{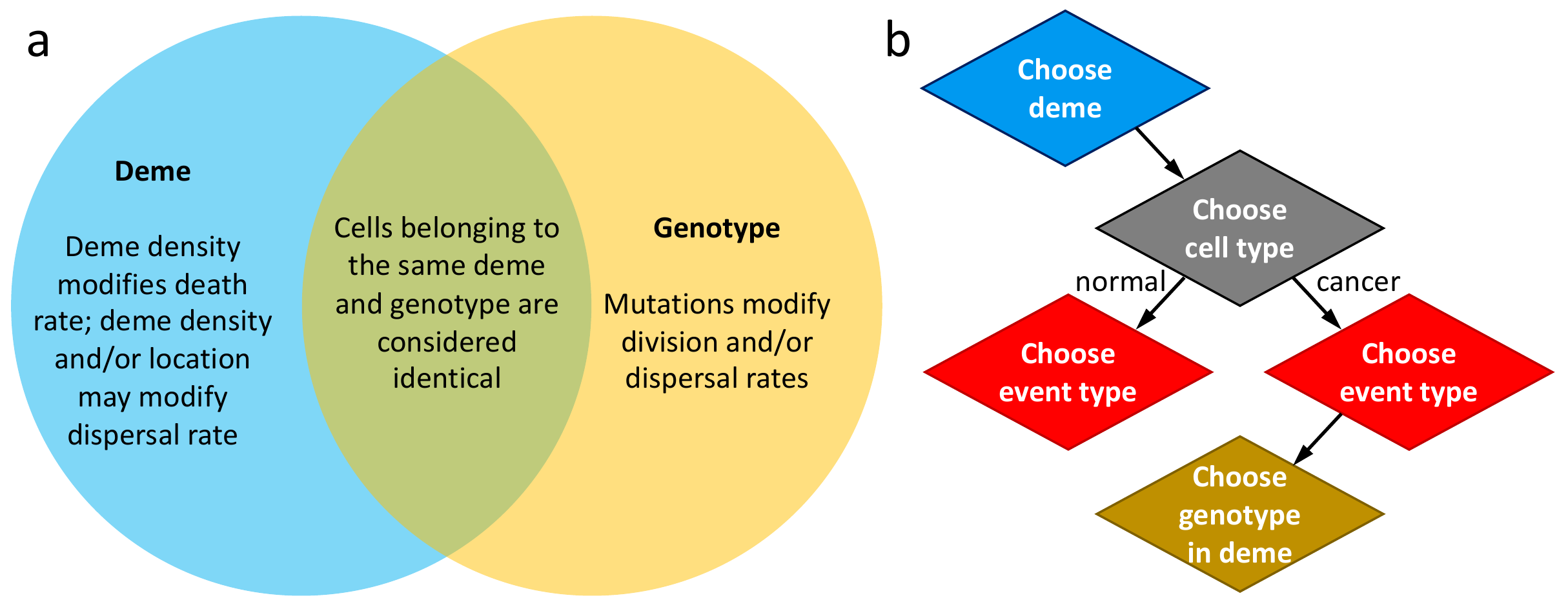}
	\caption{\textbf{a.} The basic data structures in the \textit{demon} model correspond to demes and genotypes, which are associated with population sizes and event probabilities. \textbf{b.} The program chooses between cell subpopulations and event types (cell birth, death, or dispersal) at random, with probabilities proportional to summed event rates.}
	\label{diagrams}
\end{center}
\end{figure}

For computational efficiency, the model tracks the population sizes of genotypes and demes rather than individual cells (Figure~\ref{diagrams}a). Cells that belong to the same deme-genotype intersection are considered identical. Cell events are selected sequentially using the Gillespie stochastic simulation algorithm. To determine which event to enact, the program selects at random first a deme, then a cell type (normal or tumour), thirdly an event type (cell division, death or, in the case of the tumour cell type, dispersal), and finally, for cancer cells, a genotype (Figure~\ref{diagrams}b). All these random choices are weighted by sums of event rates. Time is recorded relative to the expected cell cycle time of the initial tumour cell. 

Each random choice is made using binary search trees, so that the number of operations scales with the logarithm of the number of options. For instance, if the choice is between seven genotypes with summed event rates $r_1, \dots, r_7$ then the program will first select either set of genotypes 1-4 or set of genotypes 5-7, with probabilities $\sum_{i=1}^4 r_i \big/ \sum_{j=1}^7 r_j$ and $\sum_{i=5}^7 r_i \big/ \sum_{j=1}^7 r_j$, respectively. Supposing it chooses the first set, the program will next choose between set of genotypes 1-2 and set of genotypes 3-4. Finally, it will select between the remaining two genotypes. Sums of rates are updated after every event and are periodically recalculated from scratch to avoid rounding errors.

The \textit{demon} code is written in C++ and has been extensively tested using Valgrind\cite{nethercote_valgrind_2007}, both to ensure the absence of memory leaks and to improve computational efficiency. \textit{Demon} has been optimized to the extent that most of the execution time is spent on essential random number generation. The code is shared in a GitHub repository with instructions for installation, configuration, and execution\cite{noble_demon_2022}. Further details of the model have been described previously\cite{noble_when_2020, Noble2019b}.

\section*{Case studies}

\begin{figure}
\begin{center}
	\includegraphics[width=0.8\linewidth]{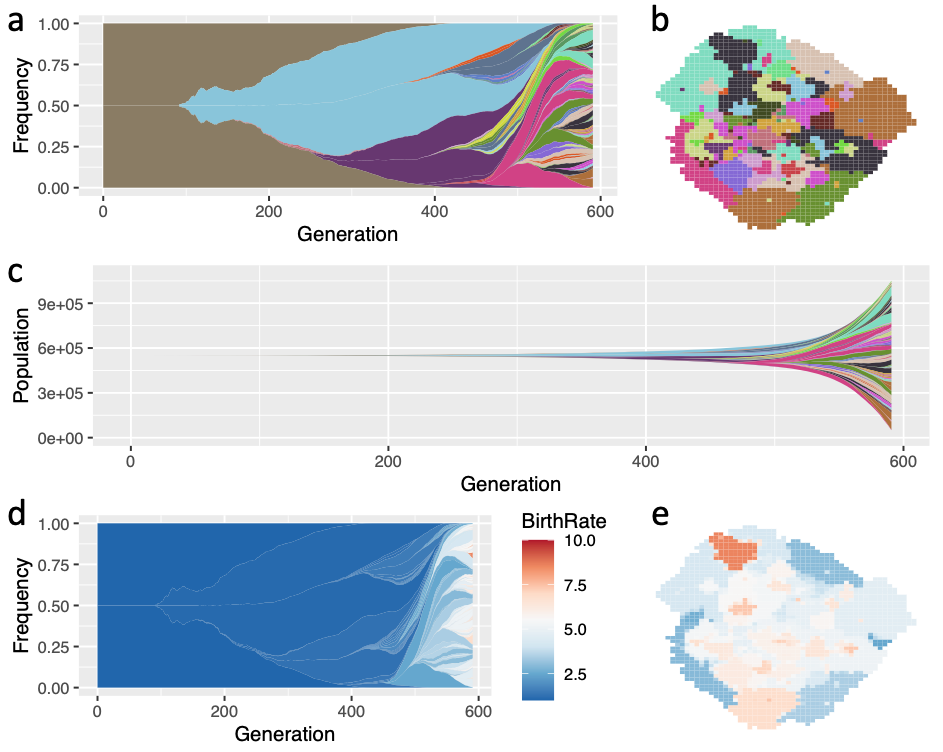}
	\caption{Visualization of tumour evolution simulated using \textit{demon}. The model in this case is a spatial Moran process, with cancer cells invading a field of normal tissue (normal cells are omitted from the plots). Cancer cells randomly acquire driver mutations that increase division rate. \textbf{a.} Muller plot in which colours represent clones with distinct combinations of driver mutations (the original clone is grey-brown; subsequent clones are coloured using a recycled palette of 26 colours). Descendant clones are shown emerging from inside their parents. \textbf{b.} Spatial plot of the tumour at the endpoint time, in which each pixel corresponds to a deme containing approximately 512 cells, coloured according to the most abundant clone within the deme. \textbf{c.} Muller plot of clone sizes, rather than frequencies. \textbf{d.} Muller plot coloured by cell division rate. \textbf{e.} Spatial plot coloured by cell division rate. This figure is adapted from reference~\cite{noble_when_2020} under the terms of a Creative Commons Attribution License, which permits use, distribution and reproduction in any medium, provided the original work is properly cited. Plots were generated using the R packages \textit{demonanalysis}\cite{noble_demonanalysis_2022} and \textit{ggmuller}\cite{noble_ggmuller_2019}.}
	\label{whenwhy}
\end{center}
\end{figure}

We have used \textit{demon} to generate results for two previous research articles. In the first of these\cite{noble_when_2020} we used the model to assess when, why and how intratumour genetic heterogeneity can be used to forecast tumour growth rate and clinical progression, thus informing the search for new prognostic biomarkers (Figure~\ref{whenwhy}). In this case we used not only the standard version of \textit{demon} but also a variant that included treatment effects and the evolution of drug resistance.

In the second study\cite{Noble2019b}, by parameterizing models with information derived from histopathological image analysis, we showed that differences in tumour architecture can explain the variety of evolutionary modes observed in human cancers. This involved running numerous model variants to discover the consequences of distinct spatial structures and to conduct sensitivity analyses. We showed that \textit{demon} output can be used to plot variant allele frequency distributions and phylogenetic trees, which can be readily compared to clinical multi-region and single-cell sequencing data.

For both of these studies we wrote custom shell and C++ scripts to run large numbers of simulations in parallel. With \textit{warlock} the HPC submission and job handling is refactored to adhere to the cluster interfaces. This makes the codebase suited for a general userbase at different computational units. 

\section*{Workflow implementation}

The \textit{warlock} repository encapsulates demon into an automated and reproducible computational workflow to simplify parallel simulations and make the software accessible to a wider community. This workflow is implemented in \textit{Snakemake}\cite{molder_sustainable_2021}, one of the most popular workflow management systems adopted by the bioinformatics community\cite{jackson_using_2021}. In addition to local execution, we use an inbuilt mechanism to provide an interface for parallel High Performance Computing via \textit{SLURM} workload manager. Hundreds of \textit{demon} instances with varied parameter values can thus be executed in parallel with a single operation. All dependencies are resolved by \textit{Anaconda}\cite{noauthor_anaconda_2020}. The analysis of model output can also be automated simply by adding further components to the pipeline.

Considering our workflow in the wider context of bioinformatics tools for computational research it seems reasonable to compare our work to Distributed Evolutionary Algorithms in Python (\textit{DEAP}). \textit{DEAP} is a well-established scientific software package designed specifically for evolutionary computations\cite{fortin_deap_2012}. It provides a general framework to implement simulations based on genetic and evolutionary algorithms. While its scope is broader and its design more abstract, our codebase is designed specifically to model tumour growth, taking into consideration parameters specific to this biological process. Moreover, \textit{DEAP} is a Python package which end users should import in their custom scripts and implement their analyses themselves, whereas our pipeline is a Snakemake workflow that executes precompiled C++ code, requiring users only to set the simulation parameters.

Our open-source software is actively developed through GitHub\cite{bak_warlock_2022-1}, making use of the platform’s Actions feature for Continuous Integration. The GitHub repository includes instructions for installation, configuration and execution. It also provides a short interactive walkthrough with example simulations and post-processing. An R package for analyzing and visualizing model output (such as in Figure~\ref{whenwhy}) continues to be developed in an additional repository\cite{noble_demonanalysis_2022}.

We have published a permanent snapshot of the first major release of our workflow on the zenodo open-access repository\cite{bak_warlock_2022}.

\bibliographystyle{unsrtnat}
{\footnotesize\bibliography{references}}

\begin{thebibliography}{24}
\providecommand{\natexlab}[1]{#1}
\providecommand{\url}[1]{\texttt{#1}}
\expandafter\ifx\csname urlstyle\endcsname\relax
  \providecommand{\doi}[1]{doi: #1}\else
  \providecommand{\doi}{doi: \begingroup \urlstyle{rm}\Url}\fi

\bibitem[Greaves and Maley(2012)]{Greaves2012}
Mel Greaves and Carlo~C Maley.
\newblock Clonal evolution in cancer.
\newblock \emph{Nature}, 481\penalty0 (7381):\penalty0 306--13, January 2012.
\newblock ISSN 1476-4687.
\newblock \doi{10.1038/nature10762}.
\newblock URL
  \url{http://www.pubmedcentral.nih.gov/articlerender.fcgi?artid=3367003&tool=pmcentrez&rendertype=abstract}.

\bibitem[Dagogo-Jack and Shaw(2018)]{Dagogo-Jack2018}
Ibiayi Dagogo-Jack and Alice~T Shaw.
\newblock Tumour heterogeneity and resistance to cancer therapies.
\newblock \emph{Nature Reviews Clinical Oncology}, 15\penalty0 (2):\penalty0
  81--94, 2018.
\newblock ISSN 17594782.
\newblock \doi{10.1038/nrclinonc.2017.166}.
\newblock URL \url{http://dx.doi.org/10.1038/nrclinonc.2017.166}.
\newblock Publisher: Nature Publishing Group ISBN: 1759-4782 (Electronic)
  1759-4774 (Linking).

\bibitem[Maley et~al.(2017)Maley, Aktipis, Graham, Sottoriva, Boddy,
  Janiszewska, Silva, Gerlinger, Yuan, Pienta, Anderson, Gatenby, Swanton,
  Posada, Wu, Schiffman, Hwang, Polyak, Anderson, Brown, Greaves, and
  Shibata]{Maley2017}
Carlo~C. Maley, Athena Aktipis, Trevor~A. Graham, Andrea Sottoriva, Amy~M.
  Boddy, Michalina Janiszewska, Ariosto~S. Silva, Marco Gerlinger, Yinyin Yuan,
  Kenneth~J. Pienta, Karen~S. Anderson, Robert Gatenby, Charles Swanton, David
  Posada, Chung-I Wu, Joshua~D. Schiffman, E.~Shelley Hwang, Kornelia Polyak,
  Alexander R.~A. Anderson, Joel~S. Brown, Mel Greaves, and Darryl Shibata.
\newblock Classifying the evolutionary and ecological features of neoplasms.
\newblock \emph{Nature Reviews Cancer}, 17\penalty0 (10):\penalty0 605--619,
  October 2017.
\newblock ISSN 1474-175X.
\newblock \doi{10.1038/nrc.2017.69}.
\newblock URL \url{http://www.nature.com/doifinder/10.1038/nrc.2017.69}.
\newblock Publisher: Nature Publishing Group.

\bibitem[Noble et~al.(2020)Noble, Burley, Le~Sueur, and
  Hochberg]{noble_when_2020}
Robert Noble, John~T. Burley, Cécile Le~Sueur, and Michael~E. Hochberg.
\newblock When, why and how tumour clonal diversity predicts survival.
\newblock \emph{Evolutionary Applications}, 13\penalty0 (7):\penalty0
  1558--1568, 2020.
\newblock ISSN 17524571.
\newblock \doi{10.1111/eva.13057}.

\bibitem[Ghaffarizadeh et~al.(2018)Ghaffarizadeh, Heiland, Friedman,
  Mumenthaler, and Macklin]{Ghaffarizadeh2018}
Ahmadreza Ghaffarizadeh, Randy Heiland, Samuel~H Friedman, Shannon~M.
  Mumenthaler, and Paul Macklin.
\newblock {PhysiCell}: {An} open source physics-based cell simulator for 3-{D}
  multicellular systems.
\newblock \emph{PLOS Computational Biology}, 14\penalty0 (2):\penalty0
  e1005991, 2018.
\newblock ISSN 1553-7358.
\newblock \doi{10.1371/journal.pcbi.1005991}.
\newblock URL \url{http://dx.plos.org/10.1371/journal.pcbi.1005991}.
\newblock ISBN: 1111111111.

\bibitem[Swat et~al.(2015)Swat, Thomas, Shirinifard, Clendenon, and
  Glazier]{Swat2015}
Maciej~H. Swat, Gilberto~L. Thomas, Abbas Shirinifard, Sherry~G. Clendenon, and
  James~a. Glazier.
\newblock Emergent {Stratification} in {Solid} {Tumors} {Selects} for {Reduced}
  {Cohesion} of {Tumor} {Cells}: {A} {Multi}-{Cell}, {Virtual}-{Tissue} {Model}
  of {Tumor} {Evolution} {Using} {CompuCell3D}.
\newblock \emph{Plos One}, 10\penalty0 (6):\penalty0 e0127972, 2015.
\newblock ISSN 1932-6203.
\newblock \doi{10.1371/journal.pone.0127972}.
\newblock URL \url{http://dx.plos.org/10.1371/journal.pone.0127972}.
\newblock ISBN: 1932-6203.

\bibitem[Bravo et~al.(2020)Bravo, Baratchart, West, Schenck, Miller, Gallaher,
  Gatenbee, Basanta, Robertson-Tessi, and Anderson]{Bravo2020}
Rafael~R. Bravo, Etienne Baratchart, Jeffrey West, Ryan~O. Schenck, Anna~K.
  Miller, Jill Gallaher, Chandler~D. Gatenbee, David Basanta, Mark
  Robertson-Tessi, and Alexander R.~A. Anderson.
\newblock Hybrid {Automata} {Library}: {A} flexible platform for hybrid
  modeling with real-time visualization.
\newblock \emph{PLOS Computational Biology}, 16\penalty0 (3):\penalty0
  e1007635, 2020.
\newblock \doi{10.1371/journal.pcbi.1007635}.
\newblock URL \url{http://dx.doi.org/10.1371/journal.pcbi.1007635}.
\newblock ISBN: 1111111111.

\bibitem[Mirams et~al.(2013)Mirams, Arthurs, Bernabeu, Bordas, Cooper, Corrias,
  Davit, Dunn, Fletcher, Harvey, Marsh, Osborne, Pathmanathan, Pitt-Francis,
  Southern, Zemzemi, and Gavaghan]{Mirams2013}
Gary~R Mirams, Christopher~J Arthurs, Miguel~O Bernabeu, Rafel Bordas, Jonathan
  Cooper, Alberto Corrias, Yohan Davit, Sara-Jane Dunn, Alexander~G Fletcher,
  Daniel~G Harvey, Megan~E Marsh, James~M Osborne, Pras Pathmanathan, Joe
  Pitt-Francis, James Southern, Nejib Zemzemi, and David~J Gavaghan.
\newblock Chaste: an open source {C}++ library for computational physiology and
  biology.
\newblock \emph{PLoS computational biology}, 9\penalty0 (3):\penalty0 e1002970,
  January 2013.
\newblock ISSN 1553-7358.
\newblock \doi{10.1371/journal.pcbi.1002970}.
\newblock URL
  \url{http://www.pubmedcentral.nih.gov/articlerender.fcgi?artid=3597547&tool=pmcentrez&rendertype=abstract}.

\bibitem[Angaroni et~al.(2022)Angaroni, Guidi, Ascolani, Onofrio, Antoniotti,
  and Graudenzi]{angaroni_j_2022}
Fabrizio Angaroni, Alessandro Guidi, Gianluca Ascolani, Alberto Onofrio, Marco
  Antoniotti, and Alex Graudenzi.
\newblock J ‑ {SPACE} : a {Julia} package for the simulation of spatial
  models of cancer evolution and of sequencing experiments.
\newblock \emph{BMC Bioinformatics}, pages 1--19, 2022.
\newblock ISSN 1471-2105.
\newblock \doi{10.1186/s12859-022-04779-8}.
\newblock URL \url{https://doi.org/10.1186/s12859-022-04779-8}.
\newblock Publisher: BioMed Central.

\bibitem[Eden(1965)]{Eden}
Murray Eden.
\newblock A {Two}-{Dimensional} {Poisson} {Growth} {Process}.
\newblock \emph{Journal of the Royal Statistical Society. Series B},
  27\penalty0 (3):\penalty0 497--504, 1965.

\bibitem[Bramson and Griffeath(1981)]{bramson_williams-bjerknes_1981}
Maury Bramson and David Griffeath.
\newblock On the {Williams}-{Bjerknes} {Tumour} {Growth} {Model} {I}.
\newblock \emph{The Annals of Probability}, 9\penalty0 (2):\penalty0 173--185,
  April 1981.
\newblock ISSN 0091-1798, 2168-894X.
\newblock \doi{10.1214/aop/1176994459}.
\newblock URL
  \url{https://projecteuclid.org/journals/annals-of-probability/volume-9/issue-2/On-the-Williams-Bjerknes-Tumour-Growth-Model-I/10.1214/aop/1176994459.full}.
\newblock Publisher: Institute of Mathematical Statistics.

\bibitem[Durrett(2015)]{Durrett2014a}
Richard Durrett.
\newblock \emph{Branching {Process} {Models} of {Cancer}}.
\newblock Springer International Publishing, Cham, 2015.
\newblock ISBN 978-3-319-16064-1.
\newblock \doi{10.1007/978-3-319-16065-8}.
\newblock URL \url{http://link.springer.com/10.1007/978-3-319-16065-8}.

\bibitem[Durrett et~al.(2016)Durrett, Foo, and Leder]{Durrett2014b}
Richard Durrett, Jasmine Foo, and Kevin Leder.
\newblock Spatial {Moran} models, {II}: cancer initiation in spatially
  structured tissue.
\newblock \emph{Journal of Mathematical Biology}, 72\penalty0 (5):\penalty0
  1369--1400, April 2016.
\newblock ISSN 0303-6812.
\newblock \doi{10.1007/s00285-015-0912-1}.
\newblock URL \url{http://link.springer.com/10.1007/s00285-015-0912-1}.

\bibitem[Nethercote and Seward(2007)]{nethercote_valgrind_2007}
Nicholas Nethercote and Julian Seward.
\newblock Valgrind: {A} {Framework} for {Heavyweight} {Dynamic} {Binary}
  {Instrumentation}.
\newblock \emph{ACM Sigplan notices}, 42\penalty0 (6):\penalty0 89--100, 2007.

\bibitem[Noble(2022{\natexlab{a}})]{noble_demon_2022}
Robert Noble.
\newblock demon, December 2022{\natexlab{a}}.
\newblock URL \url{https://github.com/robjohnnoble/demon_model}.
\newblock original-date: 2019-03-22T10:29:24Z.

\bibitem[Noble et~al.(2021)Noble, Burri, Le~Sueur, Lemant, Viossat, Kather, and
  Beerenwinkel]{Noble2019b}
Robert Noble, Dominik Burri, Cécile Le~Sueur, Jeanne Lemant, Yannick Viossat,
  Jakob~Nikolas Kather, and Niko Beerenwinkel.
\newblock Spatial structure governs the mode of tumour evolution.
\newblock \emph{Nature Ecology \& Evolution}, December 2021.
\newblock ISSN 2397-334X.
\newblock \doi{10.1038/s41559-021-01615-9}.
\newblock URL \url{https://www.nature.com/articles/s41559-021-01615-9}.

\bibitem[Noble(2022{\natexlab{b}})]{noble_demonanalysis_2022}
Robert Noble.
\newblock demonanalysis, February 2022{\natexlab{b}}.
\newblock URL \url{https://github.com/robjohnnoble/demonanalysis}.
\newblock original-date: 2018-03-22T15:51:39Z.

\bibitem[Noble(2019)]{noble_ggmuller_2019}
Robert Noble.
\newblock ggmuller: {Create} {Muller} {Plots} of {Evolutionary} {Dynamics},
  September 2019.
\newblock URL \url{https://CRAN.R-project.org/package=ggmuller}.

\bibitem[Mölder et~al.(2021)Mölder, Jablonski, Letcher, Hall, Tomkins-Tinch,
  Sochat, Forster, Lee, Twardziok, Kanitz, Wilm, Holtgrewe, Rahmann, Nahnsen,
  and Köster]{molder_sustainable_2021}
Felix Mölder, Kim~Philipp Jablonski, Brice Letcher, Michael~B. Hall,
  Christopher~H. Tomkins-Tinch, Vanessa Sochat, Jan Forster, Soohyun Lee,
  Sven~O. Twardziok, Alexander Kanitz, Andreas Wilm, Manuel Holtgrewe, Sven
  Rahmann, Sven Nahnsen, and Johannes Köster.
\newblock Sustainable data analysis with {Snakemake}.
\newblock Technical Report 10:33, F1000Research, April 2021.
\newblock URL \url{https://f1000research.com/articles/10-33}.
\newblock Type: article.

\bibitem[Jackson et~al.(2021)Jackson, Kavoussanakis, and
  Wallace]{jackson_using_2021}
Michael Jackson, Kostas Kavoussanakis, and Edward W.~J. Wallace.
\newblock Using prototyping to choose a bioinformatics workflow management
  system.
\newblock \emph{PLOS Computational Biology}, 17\penalty0 (2):\penalty0
  e1008622, February 2021.
\newblock ISSN 1553-7358.
\newblock \doi{10.1371/journal.pcbi.1008622}.
\newblock URL
  \url{https://journals.plos.org/ploscompbiol/article?id=10.1371/journal.pcbi.1008622}.
\newblock Publisher: Public Library of Science.

\bibitem[noa(2020)]{noauthor_anaconda_2020}
Anaconda {Documentation} — {Anaconda} documentation, 2020.
\newblock URL \url{https://docs.anaconda.com/}.

\bibitem[Fortin et~al.(2012)Fortin, De~Rainville, Gardner, Parizeau, and
  Gagné]{fortin_deap_2012}
Félix-Antoine Fortin, François-Michel De~Rainville, Marc-André~Gardner
  Gardner, Marc Parizeau, and Christian Gagné.
\newblock {DEAP}: evolutionary algorithms made easy.
\newblock \emph{The Journal of Machine Learning Research}, 13\penalty0
  (1):\penalty0 2171--2175, July 2012.
\newblock ISSN 1532-4435.

\bibitem[Bak(2022)]{bak_warlock_2022-1}
Maciek Bak.
\newblock warlock, November 2022.
\newblock URL \url{https://github.com/AngryMaciek/warlock}.
\newblock original-date: 2021-12-30T13:16:18Z.

\bibitem[Bak et~al.(2022)Bak, Colyer, Manojlović, and Noble]{bak_warlock_2022}
Maciej Bak, Blair Colyer, Veselin Manojlović, and Robert Noble.
\newblock Warlock: an automated computational workflow for simulating
  spatially-structured tumour evolution, December 2022.
\newblock URL \url{https://zenodo.org/record/7435093/export/hx}.
\newblock Language: eng.

\end{thebibliography}

\end{document}